\documentclass{elsart}

\usepackage{amssymb}

\begin{document}

\begin{frontmatter}



\title{Mixing of heavy baryons in the bag model calculations }


\author{A.~Bernotas\corauthref{cor}} and
\ead{bernotas@itpa.lt}
\author{V.~\v{S}imonis}
\ead{simonis@itpa.lt}
\corauth[cor]{Corresponding author.}

\address{Vilnius University Institute of Theoretical Physics and Astronomy, \newline 
A.~Go\v{s}tauto St. 12, LT-01108 Vilnius, Lithuania}

\begin{abstract}
Spin-spin interaction causes the mixing between ground state wave functions
of baryons containing three quarks of different flavors. We examine the
effect of this mixing on the baryon masses in the framework of the modified
bag model.
\end{abstract}

\end{frontmatter}

\section{Introduction}

Invented more than 30 years ago the MIT bag model \cite{01CJJTW74,02CJJT74,03DJJK75} still serves as an useful method which yields reasonable
predictions for a variety of hadronic properties (at least for the ground
states). Various aspects of the Bag model are discussed in a number of
review papers \cite{04HK78,05DD83,06T84}. Originally the model was
designed for the ultrarelativistic case of the light quarks and was rather
successful in describing the low-lying hadron spectrum. Although the first
straightforward application of the model to calculate the spectrum of the
hadrons containing heavy quarks was of very limited success because the
disagreement between calculated and observed data was evident, later on the
bag model was adjusted to incorporate the heavy quarks. It was recognized
that the reconciliation between the bag model and the heavy quark physics
could be achieved by taking into account the so-called c.m.m. (center of
mass motion) correction. For the hadrons containing only one heavy quark
there exist an approximate solution of the c.m.m. problem -- one can simply
associate the center of mass with the heavy quark and fix it \cite{07S80,08IDS82}. For the baryons containing two heavy quarks one can proceed in a
similar way assuming the heavy quarks to form the doubly-heavy diquark and
then put this object at the center of the bag \cite{09HQDLS04}. The
advantage of such approach is the simple and clear physical picture, but the
price for this simplicity is three more ore less related bag models: one for
the hadrons consisting of light quarks (in this case the role of c.m.m.
correction is partly played by the so-called zero-point energy), one for hadrons
containing single heavy quark, and one for the baryons containing two heavy
quarks. Moreover, there remains the case of the baryons consisting of three
heavy quarks which needs special treatment. Another rather popular method
to deal with c.m.m. problem is to employ a wave-packet ansatz \cite{10DJ80}.
Both approaches give only approximate solutions to the problem. However, the
second seems to be more universal and could be preferable in the case one
tries to obtain the unified description of the light and heavy hadrons. The
c.m.m. corrections is not all the story, and in order to have plausible
unified description of light and heavy hadrons (mesons and baryons) in the
framework of the bag model some other QCD inspired improvements such as
running coupling constant and running quark masses are necessary \cite
{11BS04}.

When the number of quark flavors increases we are confronted with additional
problem which needs some clarification. For the spin-$1/2$ baryons
containing three quarks of different flavors there exist two states with the
same spin and parity. We can construct the set of orthogonal wave functions
by assuming the first two quarks to be in the relative spin-$0$ or spin-$1$
state, respectively. In general case the physical states would be the linear
combinations of these mathematical states 
\begin{eqnarray}
| B\rangle
&=&C_{1} \, |(q_{1}q_{2})^{0}q_{3}\rangle+C_{2} \, |(q_{1}q_{2})^{1}q_{3}\rangle \, ,  \nonumber
\\[2ex]
| B\rangle ^{\prime }
&=&C_{2} \, |(q_{1}q_{2})^{0}q_{3}\rangle-C_{1} \, |(q_{1}q_{2})^{1}q_{3}\rangle \, .  \label{eq1.01}
\end{eqnarray}

The mixing mechanism depends on the model and approximation used. In the MIT
bag it would be the hyperfine color-magnetic interaction. The same is true
in the ordinary constituent quark model as well as in almost all variants of the
potential model. Some estimate of the state mixing is also possible in the
heavy quark effective theory \cite{12IM96}.

Because of the ambiguity in how the quarks are to be ordered in the
mathematical wave functions $|(q_{1}q_{2})^{S}q_{3}\rangle$ the expansion in Eqs.~(%
\ref{eq1.01}) is not unique. If one does not want to bother about the quark
ordering, one can simply diagonalize the interaction energy matrix \cite
{13S75}. Since such a procedure has not become a common practice, the very
natural question arises, how some authors have managed to avoid this state
mixing problem. The answer was given a long time ago in the Ref.~\cite
{14FLNC81}. The authors of that paper have shown that for the interaction
energies with quark mass dependence $\sim 1/(m_{a}m_{b})$ an optimal quark
ordering scheme could be found. The prescription is to pick the closest in
mass quarks as the first two in the wave function $|(q_{1}q_{2})^{S}q_{3}\rangle$.
Then the mixing of the states with different $S$ values is small, and the
effect of this mixing on baryon masses is negligible. Strictly speaking,
such mixing exists even in the light baryon sector between the wave functions
from which the physical states $\Sigma ^{0}$ and $\Lambda ^{0}$ are
constructed. Because of the approximate isotopic symmetry this mixing is
small, and if we are interested only in the calculation of baryon masses, we
can safely ignore this effect. The explicit calculations with the
isospin-symmetry violating terms taken into account show that the mixing
is indeed small \cite{15I80}, as expected. In the sector of charmed
baryons there are also rather strong indications that the mixing between $%
\Xi _{c}$ and $\Xi _{c}^{\prime }$ baryons is small, with negligible shifts
in the masses of these hadrons again \cite{16F97}.

And what could be said about the wave function mixing in the framework of
the bag model? Of course, we expect that all reliable models of hadron
structure yield similar results. However, we cannot apply the results of
Ref.~\cite{14FLNC81} directly because the dependence of the interaction
energy on quark masses in the bag model is somewhat more complicated. For
example, the values of light quark masses in the bag model could be set to
zero, while in the nonrelativistic models these values approach one-third
of the nucleon mass. A simple way to make the things clear is to perform
direct calculations in the bag model taking the mixing interaction into
account. This means that in the calculations of baryon energy the
off-diagonal matrix elements of the color-magnetic interaction should be
included.

In this paper we are going to examine the mixing of the ground state baryon
wave functions in the framework of the modified MIT bag model. In the next
section we give a short description of the model we are dealing with. The
concluding section contains the results of our investigation accompanied
by the discussion and some additional remarks on
the validity of approaches with and without mixing.

\section{The model}

The ground state energy of the hadron defined in the static spherical
cavity approximation is given by 
\begin{equation}
E=\frac{4\pi }{3}BR^{3}+\sum\limits_{i}n_{i}\varepsilon _{i}+\Delta E \, ,
\label{eq2.01}
\end{equation}
where $B$ is the bag constant, $R$ is the bag radius, $\varepsilon _{i}$ is 
eigenenergy of the $i$th quark in the cavity, and $\Delta E$ stands for the
interaction energy. $\Delta E$ consists of color-electric and color-magnetic
parts as described, for example, in Ref.~\cite{11BS04} in detail. For our
purpose the most important is the contribution of the color-magnetic
interaction, which in the case of the baryons containing three distinct
quarks can be written as

\begin{equation}
\Delta E_{m}=\alpha _{\mathrm{c}}(R)\sum\limits_{j>i}a_{ij}M_{ij}(m_{i},m_{j},R) \, .
\label{eq2.02}
\end{equation}
Here $\alpha _{\mathrm{c}}(R)$ is the running strong coupling constant. The function $%
M_{ij}(m_{i},m_{j},R)$ depends on quark masses and hadron bag radius and it can
be calculated explicitly. Parameters $a_{ij}$ specify the spin dependence of
the interaction energy between quarks $q_{i}$ and $q_{j}$. They are
proportional to the matrix elements $\langle(q_{1}q_{2})^{S_{1}}q_{3}|(\sigma
_{i}\cdot \sigma _{j})|(q_{1}q_{2})^{S_{2}}q_{3}\rangle$, where $\sigma _{i}$ are
appropriate spin generators. These coefficients can be calculated straightforwardly
using algebraic technique, as described in the Ref.~\cite{17C79}%
, and the transformation of the basis \cite{18YLV62}

\begin{eqnarray}
| (q_{1}q_{2})^{J_{12}}q_{3}\rangle ^{J}
&=&\sum\limits_{J_{13}}\left( -1\right) ^{j_{2}+j_{3}+J_{12}+J_{13}}\sqrt{%
\left( 2J_{12}+1\right) \left( 2J_{13}+1\right) } \\[2ex]
& & \times \left\{ 
\begin{tabular}{lll}
$j_{2}$ & $j_{1}$ & $J_{12}$ \\ 
$j_{3}$ & $J$ & $J_{13}$%
\end{tabular}
\right\}   \nonumber 
 \, | (q_{1}q_{3})^{J_{13}}q_{2}\rangle ^{J} \, ,  \label{2.03}
\end{eqnarray}
where necessary. The results are presented in the Table~\ref{t2.1}, where
for simplicity the abbreviations $|J_{12}\rangle=|(q_{1}q_{2})^{J_{12}}q_{3}\rangle$ are
used.

\begin{table}[tbp] \centering%
\caption{Parameters which specify the color-magnetic interaction energy of
baryons consisting of three distinct quarks.\label{t2.1}}%
\begin{tabular}{cccc}
\hline
& $a_{12}$ & $a_{13}$ & $a_{23}$ \\ \hline
$\langle0|a_{ij}|0\rangle$ & $-3$ &  &  \\ 
$\langle1|a_{ij}|1\rangle$ & $1$ & $-2$ & $-2$ \\ 
$\langle0|a_{ij}|1\rangle$ &  & $\sqrt{3}$ & $-\sqrt{3}$ \\ \hline
\end{tabular}
\end{table}%

The relation between the calculated bag-model energy $E$ and the hadron mass 
$M$ is given by

\begin{equation}
E=\int \mathrm{d}^{3}s \, \Phi _{P}^{2}(s)\sqrt{M^{2}+s^{2}} \, ,  \label{eq2.04}
\end{equation}
where $\Phi _{P}(s)$ is a Gauss profile 
\begin{equation}
\Phi _{P}(s)=\left( \frac{3}{2\pi P^{2}}\right) ^{3/4} \, \exp \left( -\frac{%
3s^{2}}{4P^{2}}\right) \, .  \label{eq2.05}
\end{equation}

The effective momentum $P$ specifies the momentum distribution and is
defined as

\begin{equation}
P^{2}=\gamma \sum\limits_{i}n_{i}p_{i}^{2} \, .  \label{eq2.06}
\end{equation}
Here $p_{i}$ are the momenta of the quarks. The c.m.m. parameter $%
\gamma $ is to be determined in the fitting procedure. For the baryons
containing $b$-quarks the relation ~(\ref{eq2.04}) gives practically the
same results as the familiar Einstein relation \cite{19HM90}

\begin{equation}
M^{2}=E^{2}-P^{2} \, .  \label{eq2.07}
\end{equation}
In the presence of $b$-quarks we prefer to use this simple relation
instead of rather cumbersome procedure based on Eq.~(\ref{eq2.04}). For the
running coupling constant $\alpha _{\mathrm{c}}(R)$ and running quark mass $\overline{m}%
_{f}(R)$ we use the following expressions:

\begin{equation}
\alpha _{\mathrm{c}}(R)=\frac{2\pi }{9\ln (A+R_{0}/R)} \, ,  \label{eq2.08}
\end{equation}

\begin{equation}
\overline{m}_{f}(R)=\widetilde{m}_{f}+\alpha _{\mathrm{c}}(R)\cdot \delta _{f} \, ,  \label{eq2.09}
\end{equation}
where $R_{0}$ is the scale parameter analogous to QCD constant $\Lambda $.
Parameter $A$ helps us to avoid divergences when $R\rightarrow R_{0}$. For
each quark flavor we have two free parameters $\widetilde{m}_{f}$ and $\delta
_{f}$ to be adjusted.

Let as summarize our zoo of model parameters. These are the bag constant $B$%
, the c.m.m. parameter $\gamma $ which determines the strength of the c.m.m.
corrections, two parameters ($A$ and $R_{0}$) from the running coupling
constant parametrization, and finally six parameters ($\widetilde{m}_{s}$, $%
\delta _{s}$, $\widetilde{m}_{c}$, $\delta _{c}$, $\widetilde{m}_{b}$, $\delta _{b}$%
) necessary to define the running quark mass functions $\overline{m}%
_{f}(R)$. The light (up and down) quarks are taken to be massless. To fix
the parameters $B$, $\gamma $, $A$, and $R_{0}$ the experimentally observed
masses of the light hadrons ($N$, $\Delta $, $\pi $, and the average mass of
the $\omega -\rho $ system) were chosen. To fix the mass function parameters 
$\widetilde{m}_{f}$, $\delta _{f}$ for each quark flavor we have employed the
masses of corresponding lightest vector mesons ($\phi ,$ $J/\psi ,$ $%
\Upsilon%
$) and the mass values of the lightest baryons $\Lambda _{f}$
containing the quark $q_{f}$ of the corresponding flavor. We employ the same
fitting procedure as in our previous work \cite{11BS04}, and the values of
the parameters to be used as the input in the bag model calculations are: $%
B=7.597\cdot 10^{-4}~$GeV$^{4}$, $R_{0}=2.543~$GeV$^{-1}$, $A=1.070$, $%
\gamma =1.958$, $\widetilde{m}_{s}=0.161~$GeV, $\delta _{s}=0.156~$GeV, $\widetilde{m%
}_{c}=1.458~$GeV, $\delta _{c}=0.112~$GeV, $\widetilde{m}_{b}=4.793~$GeV, $%
\delta _{b}=0.061~$GeV. The parameters $B$, $R_{0}$, $A$, $\gamma $, $\widetilde{%
m}_{s}$, $\delta _{s}$ are the same as in Ref.~\cite{11BS04}. The numerical
values of the remaining four parameters ($\widetilde{m}_{c}$, $\delta _{c}$, $\widetilde{m}_{b}$%
, $\delta _{b}$) differ slightly from the corresponding values presented in 
\cite{11BS04} because in the present work we have used new more accurate
values of $\Lambda _{c}$ (2.286$~$GeV) \cite{28AeA05} and $\Lambda _{b}$
(5.620$~$GeV) \cite{21CDF06} masses.

\section{Results and discussion}

Let us proceed to the discussion of our main point of concern -- the wave
function mixing of heavy baryons in the bag model calculations. The ground
state baryons we are interested in are $\Xi _{c}$, $\Xi _{c}^{\prime }$, $%
\Xi _{c}^{*}$; $\Xi _{b}$, $\Xi _{b}^{\prime }$, $\Xi _{b}^{*}$; $\Xi _{bc}$%
, $\Xi _{bc}^{\prime }$, $\Xi _{bc}^{*}$; and $\Omega _{bc}$, $\Omega
_{bc}^{\prime }$, $\Omega _{bc}^{*}$. The mixing is possible only between
the spin-$1/2$ states ($\Xi _{c}$ and $\Xi _{c}^{\prime }$, for example). In
order to calculate the masses of all these baryons we use the bag model
parameters listed at the end of the preceding section. For the spin-$3/2$%
{\Large \ }states denoted as $|\ldots\rangle^{*}$ the calculation procedure is exactly
the same as adopted in the paper \cite{11BS04}. We minimize the energy $%
E_{B^*}$ of each such baryon as a function of the bag radius $R$ and then
apply Eq. (\ref{eq2.04}) (for $\Xi _{c}^{*}$) or Eq. (\ref{eq2.07}) (for $%
\Xi _{b}^{*}$, $\Xi _{bc}^{*}$, and $\Omega _{bc}^{*}$) to determine the
corresponding baryon masses. For the spin-$1/2$ states $|\ldots\rangle$ and $|\ldots\rangle^{\prime }
$ the procedure differs only in the choice of the energy function to be
minimized. In this case we use the trace of the energy matrix $%
E_{B}+E_{B^{\prime }}$ which remains invariant under state mixing. Then we
calculate the diagonal and off-diagonal matrix elements of the interaction
energy, diagonalize the energy matrix, and use Eq.~(\ref{eq2.04}) or (\ref
{eq2.07}) again to determine the masses of the physical baryons. To gain
some insight how the things look like we present some intermediate results
of calculations in the Tables~\ref{t3.1}, \ref{t3.3}. In the first two rows
of these tables we give the c.m.m. uncorrected energy values $E(1)$ and $E(0)
$ corresponding to the mathematical wave functions in which the first two
quarks in the spin coupling scheme $(q_{1}q_{2})^{S}q_{3}$ are in the spin-$1
$ and spin-$0$ states. The last two rows contain the squared wave function
expansion coefficients obtained after matrix diagonalization. The symbols $b$%
, $c$, $s$ denote the bottom, charmed, and strange quarks, respectively, and
for the sake of simplicity the symbol $u$ is used for both light (up or
down) quarks.

\begin{table}[tbp] \centering%
\caption{Dependence of the calculated energies before the matrix diagonalization
 (in GeV) on the arangement of quarks for the
 $\Xi _{c}$, $\Xi _{c}^{\prime }$ (columns 2--4) and  $\Xi _{b}$, $\Xi _{b}^{\prime }$ 
(columns 5--7) baryons. The last two rows contain the squared expansion coefficients
 $C_{1}^{2}$, $C_{2}^{2}$ of the wave functions obtained after matrix 
diagonalization.\label{t3.1}}%
\begin{tabular}{lllllll}
\hline
$(q_{1}q_{2})q_{3}$ & $(us)c$ & $(uc)s$ & $(sc)u$ & $(us)b$ & $(ub)s$ & $%
(sb)u$ \\ \hline
$E(1)$ & 2.886 & 2.840 & 2.831 & 6.086 & 6.013 & 6.010 \\ 
$E(0)$ & 2.818 & 2.865 & 2.874 & 5.987 & 6.059 & 6.062 \\ 
$C_{1}^{2}$ & 0.9950 & 0.3139 & 0.1912 & 0.9997 & 0.2656 & 0.2347 \\ 
$C_{2}^{2}$ & 0.0050 & 0.6861 & 0.8088 & 0.0003 & 0.7344 & 0.7653 \\ \hline
\end{tabular}
\end{table}%

\begin{table}[tbp] \centering%
\caption{Dependence of the calculated energies before the matrix 
diagonalization (in GeV) on the arangement of quarks for the
 $\Xi _{bc}$, $\Xi _{bc}^{\prime }$ (columns 2--4)
 and $\Omega _{bc}$, $\Omega _{bc}^{\prime }$ (columns 5--7)
baryons. The last two rows contain the squared expansion coefficients
 $C_{1}^{2}$, $C_{2}^{2}$ of the wave functions obtained after matrix 
diagonalization.\label{t3.3}}%
\begin{tabular}{lllllll}
\hline
($q_{1}q_{2})q_{3}$ & $(uc)b$ & $(ub)c$ & $(cb)u$ & $(sc)b$ & $(sb)c$ & $%
(cb)s$ \\ \hline
$E(1)$ & 7.078 & 7.050 & 7.041 & 7.247 & 7.224 & 7.217 \\ 
$E(0)$ & 7.035 & 7.062 & 7.072 & 7.212 & 7.235 & 7.241 \\ 
$C_{1}^{2}$ & 0.9833 & 0.3693 & 0.1474 & 0.9872 & 0.3538 & 0.1591 \\ 
$C_{2}^{2}$ & 0.0167 & 0.6307 & 0.8526 & 0.0128 & 0.6462 & 0.8409 \\ \hline
\end{tabular}
\end{table}%

The inspection of results in Tables~\ref{t3.1} and \ref{t3.3} shows
a striking dependence of the calculated energies on the quark ordering. As
one can see, the wave function with two first quarks in the relative spin-$0$
state has the lowermost energy only when the heaviest quark (e.\,g., $b$-quark)
is picked up as the third in the corresponding spin coupling scheme $%
(q_{1}q_{2})^{S}q_{3}$. This is the only case when the traditional
prescription 
\begin{equation}
|\ldots\rangle=|(q_{1}q_{2})^{0}q_{3}\rangle \, , \qquad |\ldots\rangle^{\prime }=|(q_{1}q_{2})^{1}q_{3}\rangle \, ,
\label{eq3.01}
\end{equation}
could be maintained, because the energy matrix diagonalization leads to
negligible changes of the initial energy values. The direct calculations
show that even in the most problematic case of the $\Xi _{bc}-\Xi
_{bc}^{\prime }$ system the difference between energy values before and
after diagonalization does not exceed 1 MeV and is obviously much smaller than
the systematic uncertainties of the model. So, if one is interested only in
the baryon mass spectra, one can adopt the prescription (\ref{eq3.01}),
construct the optimal basis by arranging the quarks in increasing order of
their masses, and not to bother about the diagonalization of the energy
matrix anymore. At this point a remark is necessary. One must be
very cautious when dealing with other baryon parameters (such as magnetic
moments, for example). As it was shown in Ref.~\cite{14FLNC81}, the wave
function mixing may change the values of the calculated magnetic moments
substantially even when the optimal basis is used. Although this problem is
beyond the scope of the present paper, it is worth attention, and we are
going to return to this question in the future.

Before going to the concluding remarks we want to compare the masses of
baryons calculated in our work with the results obtained in other models and
experimental data where available. We have chosen for the sake of comparison
the baryon mass estimates in nonrelativistic \cite{22KLPS02} and
relativistic \cite{23EFG05,24EFGM02} potential models obtained in the
quark-diquark approximation, the estimates obtained in the quark-diquark
approximation of the bag model \cite{09HQDLS04}, the calculations in the
simplified variational approach \cite{25AAHN05}, and predictions provided
using various sum rules based partially on the heavy quark symmetry
considerations \cite{26LRP96,27RLP95}. The experimental values are taken
from the Particle Data Tables \cite{20PDG06}. The data for the baryons of $%
\Xi _{Q}$ type are presented in the Table~\ref{t3.5} and for the $\Xi
_{Q_{1}Q_{2}}$, $\Omega _{Q_{1}Q_{2}}$ type baryons in the Table~\ref{t3.6}.

\begin{table}[tbp] \centering%
\caption{Masses of $\Xi _{c}$, $\Xi _{c}^{\prime }$, $\Xi _{c}^{*}$ and
 $\Xi _{b}$, $\Xi _{b}^{\prime }$, $\Xi _{b}^{*}$ baryons
 (in GeV). The row denoted as Bag contains the results obtained in our work.
 The row Exp contains averaged over the isodoublet experimental energy values.\label{t3.5}}%
\begin{tabular}{lllllll}
\hline
Particle & $\Xi _{c}$ & $\Xi _{c}^{\prime }$ & $\Xi _{c}^{*}$ & $\Xi _{b}$ & 
$\Xi _{b}^{\prime }$ & $\Xi _{b}^{*}$ \\ \hline
Bag & 2.468 & 2.546 & 2.638 & 5.809 & 5.911 & 5.944 \\ 
\cite{23EFG05} & 2.481 & 2.578 & 2.654 & 5.812 & 5.937 & 5.963 \\ 
\cite{25AAHN05} & 2.474 & 2.578 & 2.655 & 5.808 & 5.946 & 5.975 \\ 
\cite{26LRP96,27RLP95} & 2.468 & 2.582 & 2.651 & 5.810 & 5.955 & 5.984 \\ 
Exp & 2.469 & 2.577 & 2.646 & -- & -- & -- \\ \hline
\end{tabular}
\end{table}%

\begin{table}[tbp] \centering%
\caption{Masses of $\Xi _{bc}$, $\Xi _{bc}^{\prime }$, $\Xi _{bc}^{*}$ and
 $\Omega _{bc}$, $\Omega _{bc}^{\prime }$, $\Omega _{bc}^{*}$ baryons (in GeV). 
The row denoted as Bag contains the results obtained in our work.\label{t3.6}}
\begin{tabular}{lllllll}
\hline
Particle & $\Xi _{bc}$ & $\Xi _{bc}^{\prime }$ & $\Xi _{bc}^{*}$ & $\Omega
_{bc}$ & $\Omega _{bc}^{\prime }$ & $\Omega _{bc}^{*}$ \\ \hline
Bag & 6.846 & 6.891 & 6.919 & 6.999 & 7.036 & 7.063 \\ 
\cite{22KLPS02} & 6.82 & 6.85 & 6.90 & 6.93 & 6.97 & 7.00 \\ 
\cite{24EFGM02} & 6.933 & 6.963 & 6.980 & 7.088 & 7.116 & 7.130 \\ 
\cite{26LRP96} & 7.029 & 7.053 & 7.083 & 7.126 & 7.148 & 7.165 \\ 
\cite{09HQDLS04} & 6.838 & 7.028 & 6.989 & 6.941 & 7.116 & 7.077 \\ \hline
\end{tabular}
\end{table}%

From Table~\ref{t3.5} it is seen that for the baryons containing one heavy
quark all approaches give rather similar qualitative picture. Inspection of
the $\Xi _{Q}~-~\Xi _{Q}^{*}$ hyperfine mass splitting indicates that in our
version of the bag model the interaction energies for these baryons could be
slightly underestimated. Comparison with experiment also shows that all
approaches give reasonable results. One could even insist that owing to the
approximate nature of the models the agreement with experiment (though not
excellent) is surprisingly good. Such success gives us some confidence that
we are on the right path in understanding the properties of heavy baryons.

For the baryons with two heavy quarks the situation is somewhat different.
As seen from Table~\ref{t3.6}, all but one approaches give similar
qualitative pictures of the baryon spectra again. A striking exception is
the results obtained in the paper \cite{09HQDLS04} (the reversed order of
the $\Xi _{bc}^{\prime }$, $\Xi _{bc}^{*}$ and $\Omega _{bc}^{\prime }$, $%
\Omega _{bc}^{*}$ states). The bag model results for the ground state baryon
masses calculated in our work are laid out somewhat above the estimates \cite
{22KLPS02} obtained in the nonrelativistic potential model based on the
quark-diquark approximation. Relativistic approach \cite{24EFGM02} gives
similar mass spectrum as ours but shifted approximately 70 MeV upwards. The
predictions based on the sum rules \cite{26LRP96} are higher than our
estimates by approximately 170 MeV and 110 MeV for the $\Xi _{bc}$ and $%
\Omega _{bc}$ families respectively. The difference between the baryon
spectrum obtained in the paper \cite{09HQDLS04} and the others is of
qualitative character. It could look strange, but it is the direct
consequence of the attempts to incorporate the mixing effects for the ground
state baryons in the quark-diquark approximation to the bag model. We
already know that in the ordinary approach the wave function mixing can play
an important role in the calculations of the baryon mass spectra. However,
in general, we cannot draw a direct link between the quark ordering in
the spin coupling scheme and the corresponding diquark structure.
Nevertheless, some correspondence between the two pictures is expected. 
For example, in the quark-diquark approximation to the potential
model the physical $\Xi _{bc}^{\prime }$ and $\Omega _{bc}^{\prime }$ states
are those with scalar $cb$ diquark \cite{24EFGM02}, as could be expected
from the analogy with the ordinary approach (see the 4th and 7th
columns of Table~\ref{t3.3}). In the usual approach the interaction of
the system consisting of two quarks with the third one is provided by
the interaction of its individual constituents. The mixing of the wave
functions is possible only when the interaction between the first and the 
third quarks is of different strength as compared with the interaction between 
the second and the third (as the quark becomes heavier its hyperfine interaction with other quarks
decreases). When the mixing is present the correct mass splitting is
achieved only after the diagonalization of the energy matrix. On the other
hand, in the quark-diquark approximation some information about the initial
structure of the diquark is lost, and, as a rule, the mixing of the ground state
functions is absent \cite{23EFG05}. Maybe some remnant of the mixing
interaction of the ground states could exist, but practically it seems to be
not necessary. Since the baryon masses predicted in the paper 
\cite{09HQDLS04} differ radically from the predictions obtained in the bag
model with the state mixing effects taken into account (this work) and from
the results obtained in the quark-diquark approximations to the potential
model, it seems that the mixing effects in the work \cite{09HQDLS04} have
been heavily overestimated. Of course, in the calculation of energies of
the excited baryons one is confronted with the mixing of various states, and
in consistent calculations \cite{22KLPS02,24EFGM02} these mixing effects
are taken into account.

As regards the results obtained in our work, first of all, we conclude that,
as expected, the bag model shares many features of ordinary quark model. The
main aim of this paper was to examine the heavy baryon ground state wave
function mixing due to the color-magnetic interaction in the framework of
the modified bag model. We have found that the main features of the mixing
interaction in the bag model are the same as in the ordinary nonrelativistic
quark model. So, we can conclude that fully relativistic treatment of the
light quarks in the bag has only minor influence on the state mixing
properties. For the baryons consisting of three quarks of different flavor
we cannot in general ignore the wave function mixing induced by the hyperfine
color-magnetic interaction. It can even cause sizable changes of the
calculated hadronic properties. On the other hand, the widely accepted
optimal basis can be built up by simply choosing the heaviest quark as the
third one in the corresponding spin coupling scheme. The matrix of the
interaction energy in this basis is approximately diagonal, and therefore the
mixing effects in the baryon mass (energy) calculations can be neglected. If
for any reason other than optimal basis is used, even in the baryon mass
calculations the mixing effects must be taken into account.


\begin{thebibliography}{00}

\bibitem{01CJJTW74}
A. Chodos, R.L. Jaffe, K. Johnson, C.B. Thorn, V.F.~Weisskopf, Phys. Rev.~D~9 (1974) 3471.
\bibitem{02CJJT74}
A. Chodos, R.L. Jaffe, K. Johnson, C.B. Thorn, Phys. Rev.~D~10 (1974) 2599.
\bibitem{03DJJK75}
T. DeGrand, R.L. Jaffe, K. Johnson, J. Kiskis, Phys. Rev.~D~12 (1975) 2060.
\bibitem{04HK78}
P. Hasenfrantz, J. Kuti, Phys. Rep. 40 (1978) 75.
\bibitem{05DD83}
C.E. DeTar, J.F. Donoghue, Ann. Rev. Nucl. Part. Sci. 33 (1983) 238.
\bibitem{06T84}
A.W. Thomas, Adv. Nucl. Phys. 13 (1984) 1.
\bibitem{07S80}
E.V. Shuryak, Phys. Lett. B 93 (1980) 134.
\bibitem{08IDS82}
D. Izatt, C. DeTar, M. Stephenson, Nucl. Phys. B 199 (1982) 269.
\bibitem{09HQDLS04}
D. He, K. Qian , Y. Ding, X. Li, P. Shen, Phys. Rev.~D~70 (2004) 094004.
\bibitem{10DJ80}
J.F. Donoghue, K. Johnson, Phys. Rev. D 21 (1980) 1975.
\bibitem{11BS04}
A. Bernotas, V. \v{S}imonis, Nucl. Phys. A. 741 (2004) 179.
\bibitem{12IM96}
T. Ito, Y. Matsui, Prog. Theor. Phys. 96 (1996) 659.
\bibitem{13S75}
D. Sakharov, Pis'ma Zh. Eksp. Teor. Fiz. 21 (1975) 554 [JETP Lett. 21 (1975)~9].
\bibitem{14FLNC81}
J. Franklin, D.B. Lichtenberg, W. Namgung, D. Carydas, Phys. Rev.~D~24 (1981) 2910.
\bibitem{15I80}
N. Isgur, Phys. Rev. D 21 (1980) 779.
\bibitem{16F97}
J. Franklin, Phys. Rev. D 55 (1997) 425.
\bibitem{17C79}
F.E. Close, An Introduction to Quarks and Partons (Academic Press, 1979).
\bibitem{18YLV62}
A.P. Yutsis, I.B. Levinson, V.V. Vanagas, The Theory of Angular Momentum (Israel Program for Scientific Translations, Jerusalem, 1962).
\bibitem{19HM90}
L.C.L. Hollenberg, B.H.J. McKellar, J. Phys. G 16 (1990) 31.
\bibitem{28AeA05}
B. Aubert et al. (\textsl{BABAR} Collaboration), Phys. Rev. D 72 (2005) 052006.
\bibitem{21CDF06}
D. Acosta et al. (CDF Collaboration), Phys. Rev. Lett. 96 (2006) 202001.
\bibitem{22KLPS02}
V.V. Kiselev, A.K. Likhoded, O.N. Pakhomova, V.A. Saleev, Phys. Rev.~D~66 (2002) 034030.
\bibitem{23EFG05}
D. Ebert, R.N. Faustov, V.O. Galkin, Phys. Rev. D 72 (2005) 034026.
\bibitem{24EFGM02}
D. Ebert, R.N. Faustov, V.O. Galkin, A.P. Martynenko, Phys. Rev.~D~66 (2002) 014008.
\bibitem{25AAHN05}
C. Albertus, J.E. Amaro, E. Hernandez, J. Nieves, Nucl. Phys.~A~755 (2005) 439.
\bibitem{26LRP96}
D.B. Lichtenberg, R. Roncaglia, E. Predazzi, Phys. Rev.~D 53 (1996) 6678.
\bibitem{27RLP95}
R. Roncaglia, D.B. Lichtenberg, E. Predazzi, Phys. Rev.~D 52 (1995) 1722.
\bibitem{20PDG06}
Particle Data Group (W.-M. Yao et al.), J. Phys. G 33 (2006) 1.
\end{thebibliography}
\end{document}